\documentclass[12pt]{iopart}
\pdfoutput=1 
\usepackage[pdftex]{graphicx,color}
\usepackage{amssymb}
\usepackage{epsfig}
\usepackage{subfigure}
\usepackage{textcomp}
\usepackage{cite}
\usepackage{float}
\usepackage{bm}
\expandafter\let\csname equation*\endcsname\relax
\expandafter\let\csname endequation*\endcsname\relax
\usepackage{amsmath}
\pdfoutput=1 
\usepackage{amsmath,amssymb,eucal,graphicx,float,epstopdf}
\graphicspath{ {./figs/} }

\begin{document}
\title{Emergent Network Modularity}

\author{P. L. Krapivsky} \address{Department of Physics, Boston University,
  Boston MA 02215, USA}
  
\author{S. Redner} \address{Santa Fe Institute, 1399 Hyde Park Road, Santa
  Fe, New Mexico 87501, USA}

\begin{abstract}

  We introduce a network growth model based on complete redirection: a new
  node randomly selects an existing target node, but attaches to a random
  neighbor of this target.  For undirected networks, this simple growth rule
  generates unusual, highly modular networks.  Individual network realizations
  typically contain multiple macrohubs---nodes whose degree scales linearly
  with the number of nodes $N$.  The size of the network ``nucleus''---the
  set of nodes of degree greater than one---grows sublinearly with $N$ and
  thus constitutes a vanishingly small fraction of the network.  The network
  therefore consists almost entirely of leaves (nodes of degree one) as
  $N\to\infty$.

\end{abstract}


\section{Introduction}

\begin{figure}[ht]
\centerline{\subfigure[]{\includegraphics[width=0.275\textwidth]{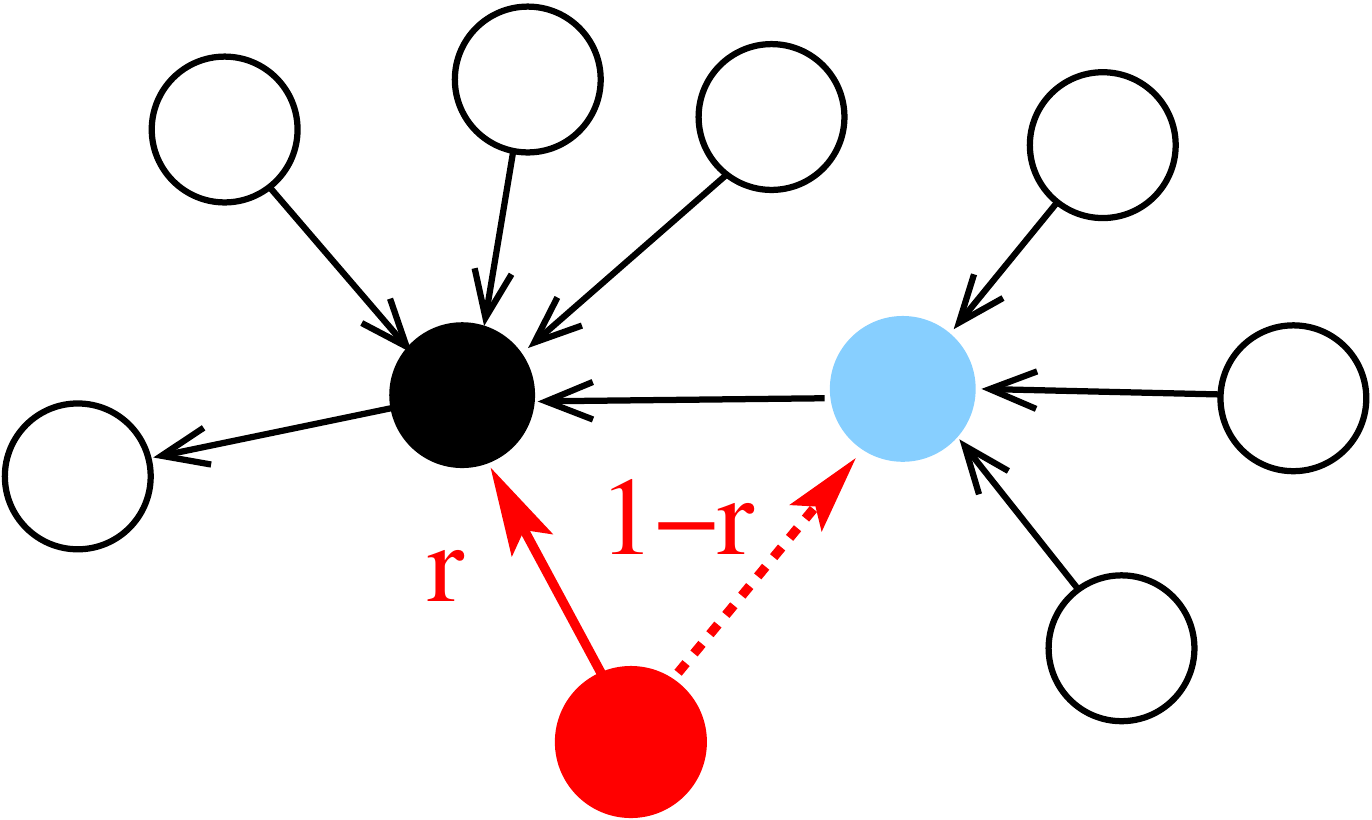}}\qquad\qquad
\subfigure[]{\includegraphics[width=0.275\textwidth]{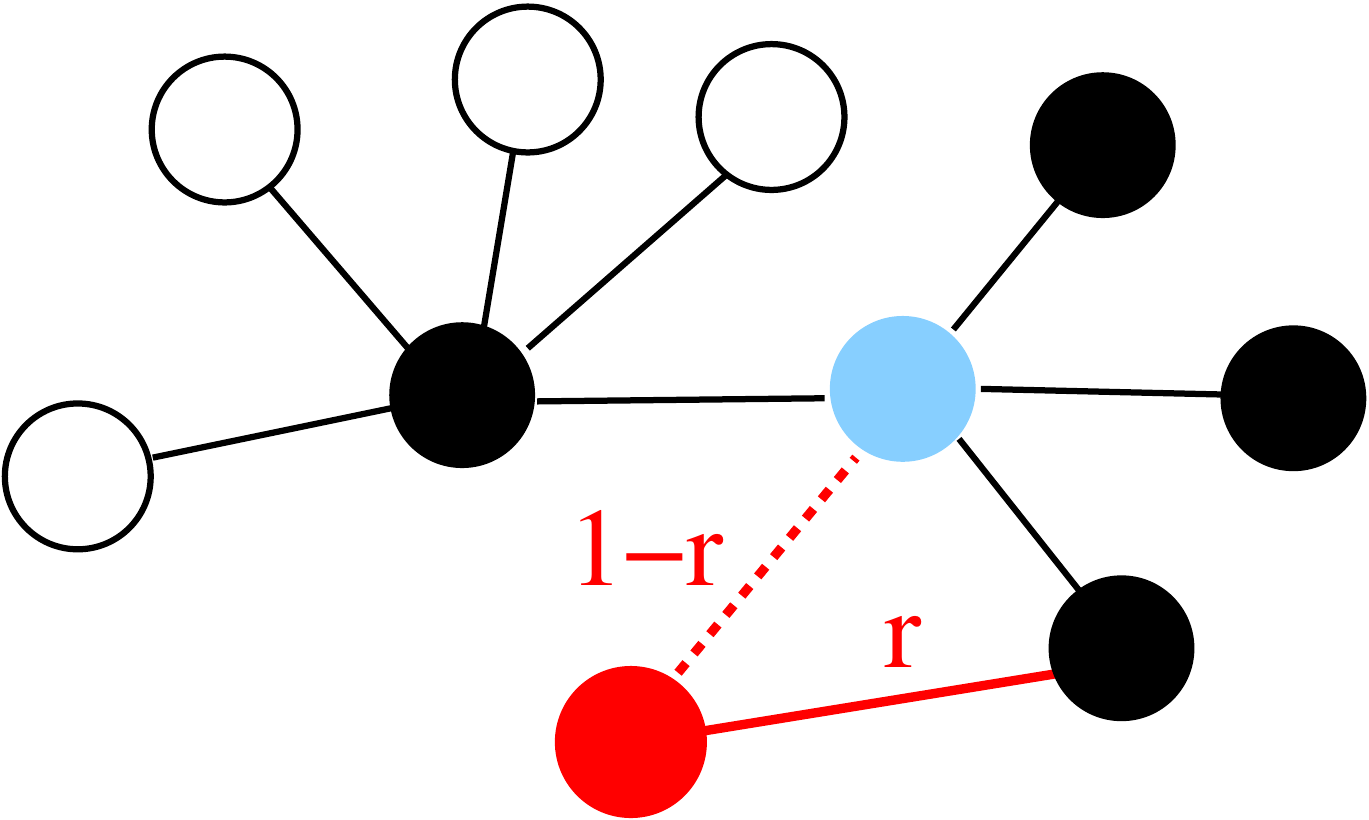}}}
\caption{Redirection for (a) directed and (b) undirected networks.
  (a) The new node attaches by redirection to the unique ancestor (black) of
  the provisional target (light blue).  (b) With the same target, the new
  node attaches to one of the black nodes.}
\label{model}
\end{figure}

Redirection is a fundamental network growth mechanism to determine how a new
node attaches to a growing network.  For \emph{directed} networks, with a
prescribed direction for each link, redirection is implemented as follows
(Fig.~\ref{model}(a)):
\begin{enumerate}
\setlength\itemsep{-0.35ex}
\item A new node chooses a provisional target node uniformly at random.
\item With probability $0\leq 1-r\leq 1$, the new node attaches to this
  target.
\item With probability $r$, the new node attaches to the ancestor of the
  target.
\end{enumerate}
By its very construction, an initial tree network always remains a tree.

Without the redirection step (iii), the above growth rules define the random
recursive tree (RRT)~\cite{NR70,M74,MM78}, for which the average number of
nodes of degree $k$, $N_k= N/2^k$, where $N$ is the total number of nodes.
Redirection represents a minimalist extension of the RRT; this idea was
suggested in~\cite{kum} and was made more concrete and developed
mathematically in~\cite{KR01}.  The latter work showed that redirection is
equivalent to \emph{shifted linear preferential attachment}, in which the
rate of attaching to a pre-existing network node of degree $k$ is
proportional to $k+\lambda$, with $\lambda =\frac{1}{r}-2$.  That is,
redirection transforms a purely local growth mechanism into a global
mechanism.  Redirection is also extremely efficient algorithmically; to build
a network of $N$ nodes requires a computation time that scales linearly with
$N$, with a prefactor of the order of one.

Because of its useful qualities, redirection has been extended in many ways:
connecting to (i) more distant ancestors~\cite{V03}; (ii) an arbitrary
ancestor~\cite{BK}; (iii) all earlier ancestors~\cite{KR_log}; (iv) multiple
ancestors~\cite{RA04}; and (v) the ancestor with a probability that depends
on the degrees of the provisional target and the ancestor~\cite{GR13,GKR13}.
Each of these scenarios has revealed intriguing features that highlight the
richness of the redirection mechanism.

Although this growth mechanism has been applied to directed networks,
\emph{undirected} graphs are more pertinent for many applications.  In social
networks, for example, directionality plays a limited role because friendship
is inherently a two-way relationship~\cite{WF94}.  This observation motivates
us to extend redirection to \emph{undirected} networks.  Such an isotropic
network again grows according to the rules enumerated above, but now
redirection can occur to \emph{any} of the neighbors of the provisional
target (Fig.~\ref{model}(b)).  We define this process as \emph{isotropic
  redirection} (IR).  While the behavior of this IR model for general
redirection probability $0<r< 1$ is interesting in its own
right~\cite{FSRA14,AKR18}, here we focus on the parameter-free case of $r=1$,
where the new node always attaches to a random neighbor of the provisional
target.

\begin{figure}[ht]
\centerline{\raisebox{0.0cm}{\includegraphics[width=0.32\textwidth]{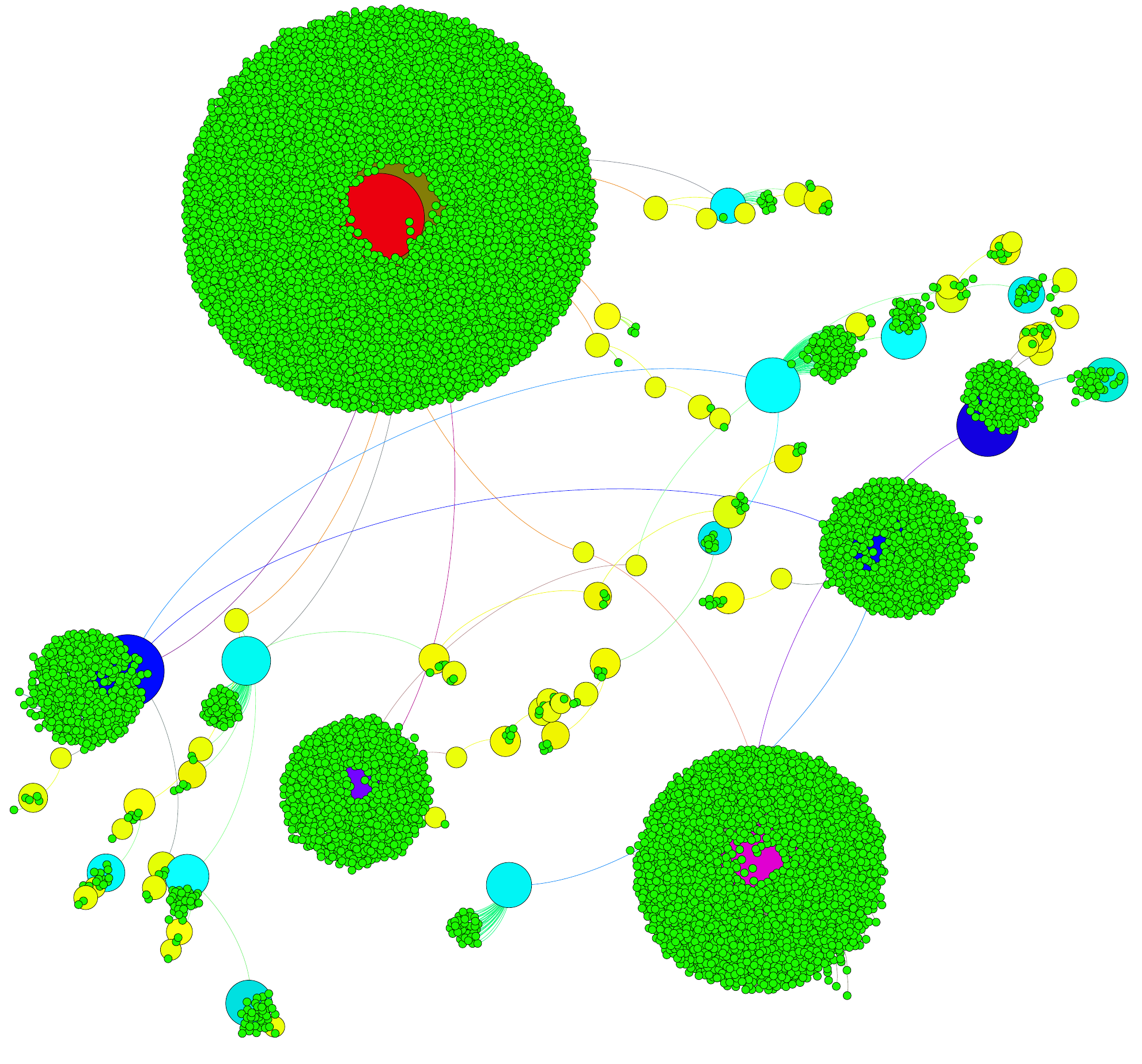}}
\raisebox{0.0cm}{\includegraphics[width=0.32\textwidth]{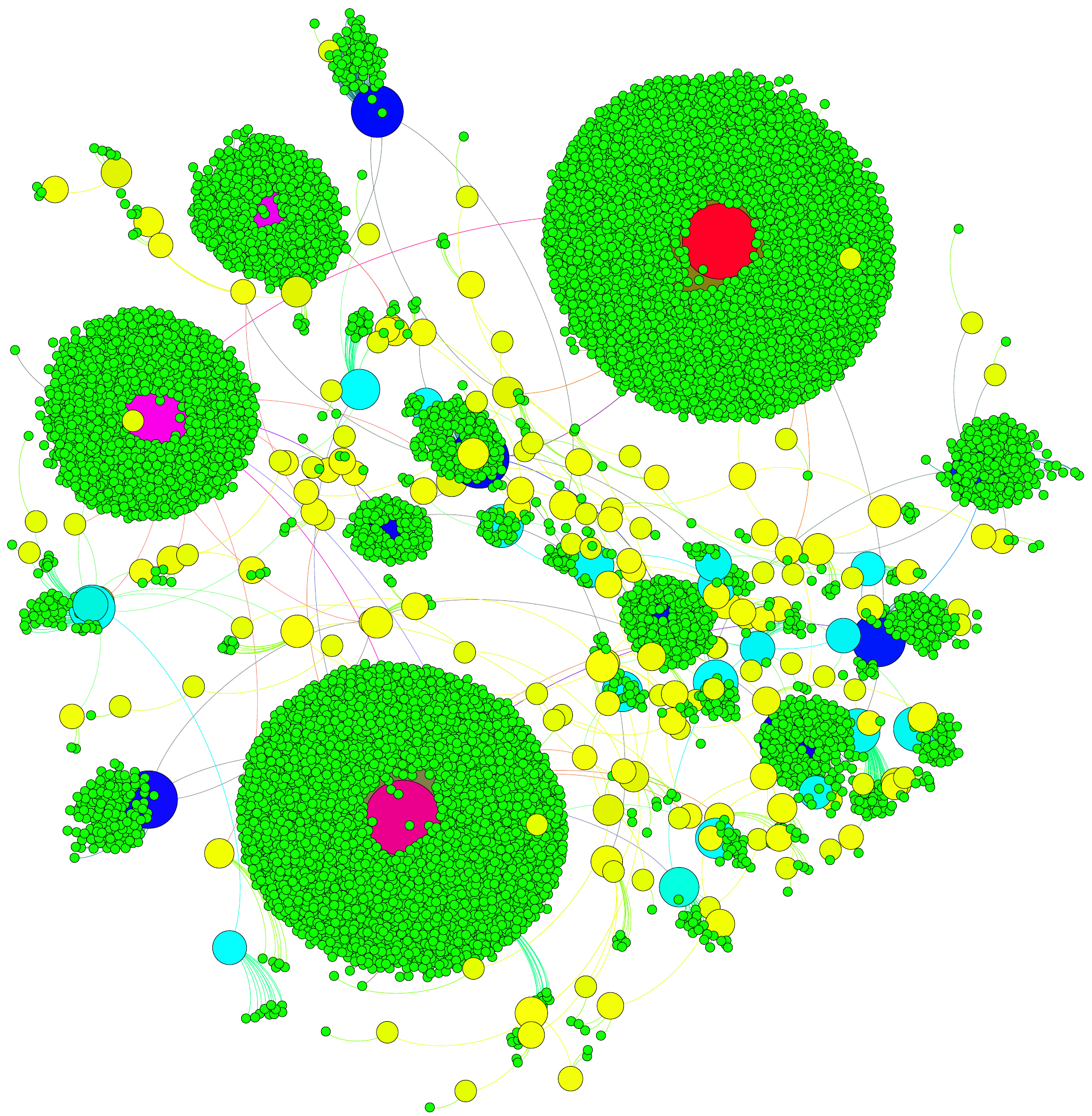}}
\raisebox{-0.0cm}{\includegraphics[width=0.32\textwidth]{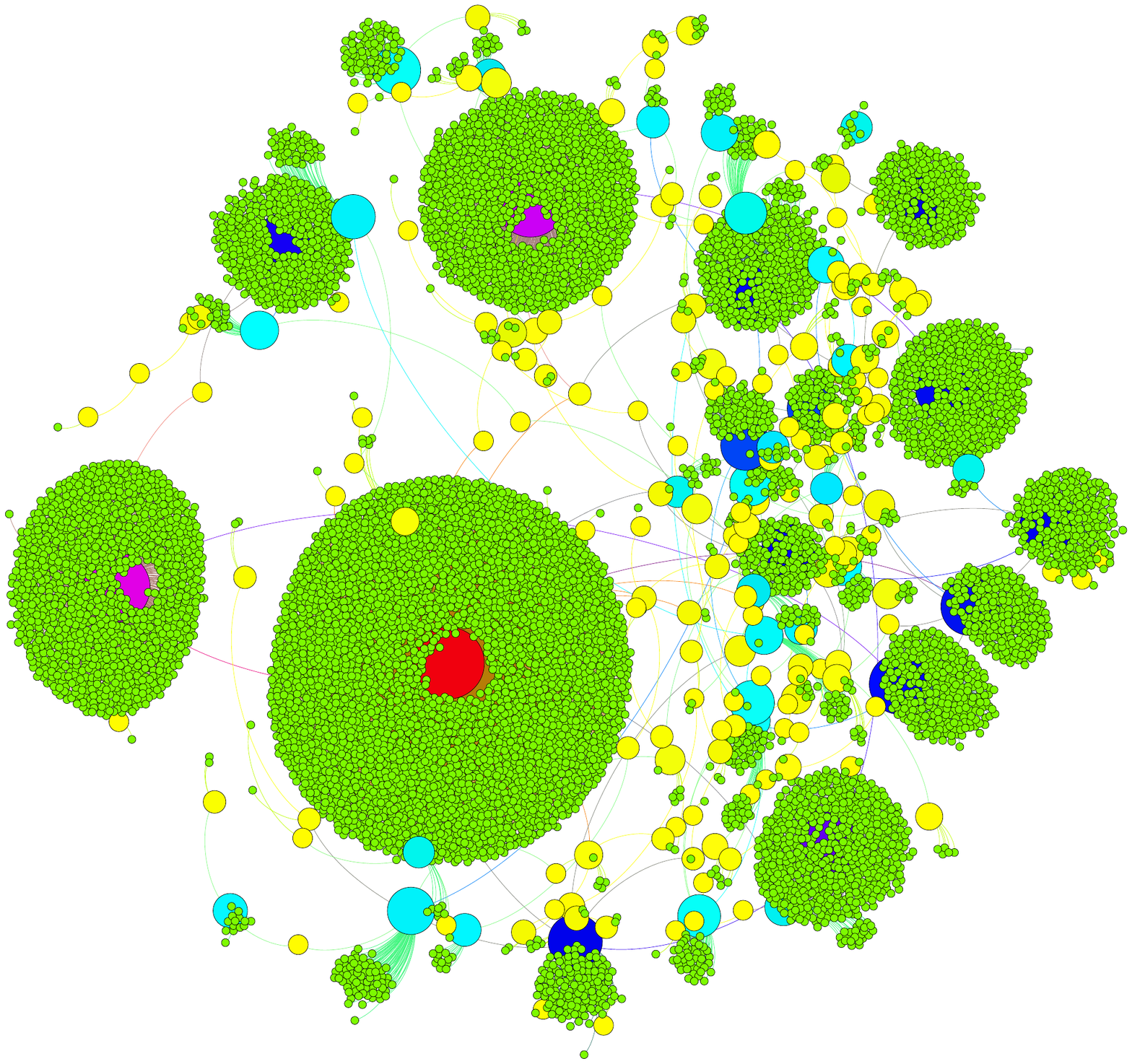}}}
\caption{ Examples of networks of $10^4$ nodes grown by redirection.  Green:
  nodes of degree $k=1$; yellow, $2\!\leq\! k\!\leq\! 10$; cyan,
  $11\!\leq\! k\!\leq\! 99$; blue $100\!\leq\! k\!\leq\! 500$; violet $\to$
  red, $k\!>\!501$.  The node radius also indicates its degree. }
\label{examples}
\end{figure}

The consequences of this IR growth rule are surprisingly profound, as highly
modular networks emerge (Fig.~\ref{examples}).  Typical network realizations
contain a number of well-resolved modules, each with a central macrohub whose
degree is a finite fraction of the total number of nodes $N$.  These modules
visually resemble a variety of multiplex, or multilayer,
networks~\cite{BPPSH10,BBCG14,KABGMP14,AS14}.  Typical networks also consist
almost entirely of leaves (nodes of degree 1) as $N\to \infty$; that is, the
number of leaves satisfies $N_1/N\to 1$ as $N\to\infty$.  Nodes with degrees
$k>1$ constitute what we term the ``nucleus'' of the network.  This nucleus
comprises an infinitesimal fraction of the network, as the number of nucleus
nodes $\mathcal{N}= \sum_{k\ge 2}N_k$ grows as $N^{\mu}$, with
$\mu\approx 0.566$.  

The number of nodes of degree $k$ grows in a similar manner:
$N_k\sim N^{\mu}$ for any $k\geq 2$, with an algebraic tail
\begin{equation}
\label{Nk:large}
N_k\sim \frac{N^\mu}{k^{1+\mu}}
\end{equation} 
when $k\gg 1$.  Thus the degree distribution grows sublinearly with network
size ($\mu<1$), so that the degree exponent $1+\mu$ is generally less than
2.  These features strongly contrast with known sparse networks, where the
nucleus and the degree distribution grows linearly with network size (see,
e.g., \cite{N10}) and has the algebraic tail
\begin{align*}
N_k\sim \frac{N}{k^{\nu}}\,.
\end{align*}
Here, the degree exponent satisfies $\nu>2$ and depends on the model (see
\cite{N10}), while in the IR model (as well as in the models of
Ref.~\cite{GKR13}) the growth exponent $\mu$ fixes the degree exponent to be
$1+\mu$, which ensures that it is less than 2.  Finally, we emphasize that
for directed networks, redirection with $r=1$ generates a star graph; hence
bidirectional links are needed to generate non-trivial networks.

In Sec.~\ref{sec:star}, we begin by showing that star-like structures are
surprisingly common in IR networks.  We investigate the probability
distribution for maximal degrees in Sec.~\ref{sec:MD}.  We then show that
multiple macrohubs arise with an anomalously large probability in
Sec.~\ref{sec:H}.  In Sec.~\ref{sec:core}, we discuss basic features of the
nucleus of the network.  Finally in Sec.~\ref{sec:book} we study the
intriguing features that arise when a new node attaches to multiple neighbors
of the provisional target.

\section{Perfect and Near-Perfect Star Graphs}
\label{sec:star}

\begin{figure}[ht]
\centerline{\includegraphics[width=0.9\textwidth]{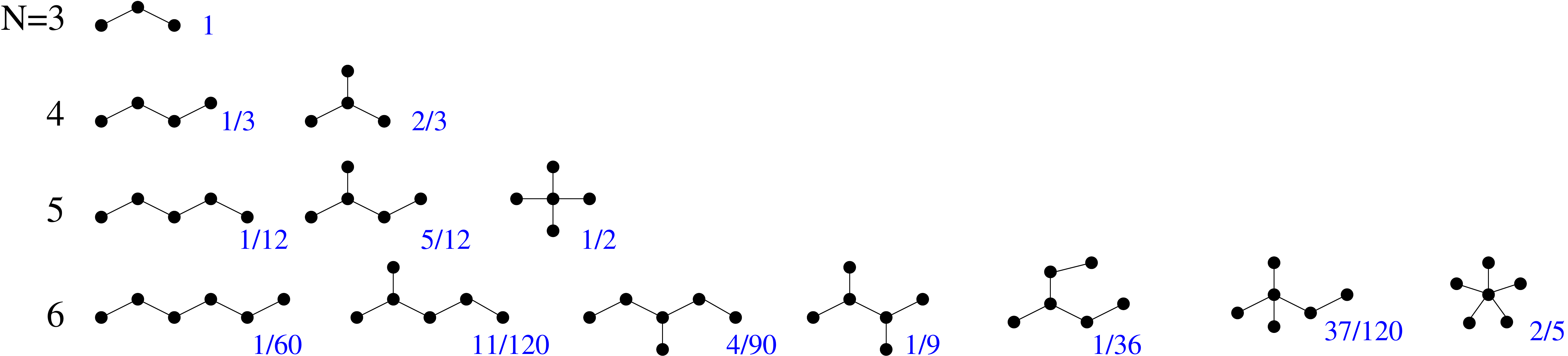}}
\caption{Enumeration of all network configurations up to $N=6$ nodes. }
\label{enum}
\end{figure}

Unless otherwise stated, we assume that the initial network is a dimer:
$\bullet$\!\rule[0.5ex]{0.35cm}{0.55pt}\!$\bullet$.  For $N=3$, there is a
single unique graph.  For $N=4$, a star occurs with probability $\frac{2}{3}$
by the new node selecting either of the leaves of the 3-node graph, after
which redirection leads to attachment to the central node.  Conversely, a
linear chain is created with probability $\frac{1}{3}$.  All IR network
realizations of up to 6 nodes and their weights are shown in Fig.~\ref{enum}.

\begin{figure}[ht]
\centerline{\includegraphics[width=0.7\textwidth]{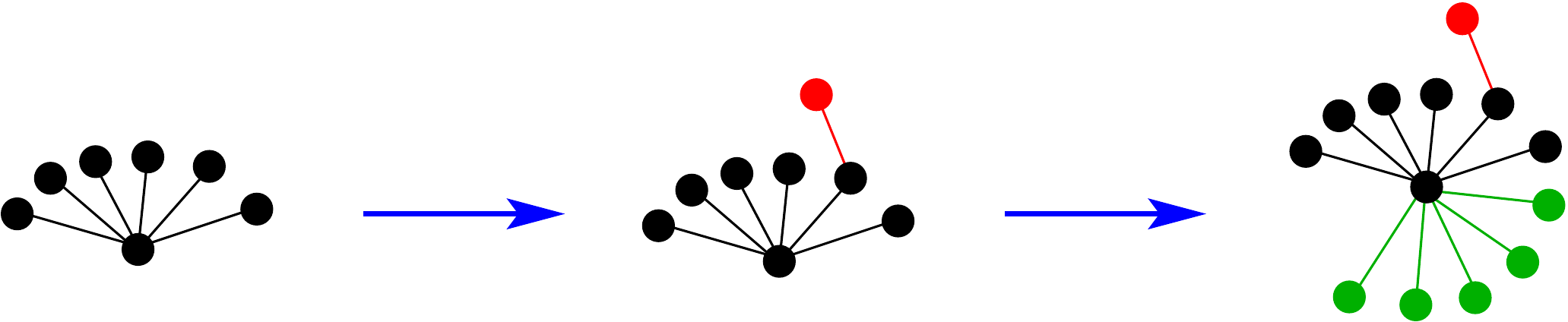}}
\caption{A single-defect star.  A perfect star (black) is built to an
  intermediate stage, then an error occurs (red).  All subsequent attachments
  (green) are to the hub.}
\label{defect-star}
\end{figure}

While it is impractical to extend this enumeration to large $N$, we can
compute the probabilities to generate the special configurations of a perfect
star and near-perfect stars of $N$ nodes.  Let $S_N$ be the probability to
create a perfect star of $N$ nodes.  To build this star, a new node has to
provisionally select one of the periphery nodes (which occurs with
probability $(k-1)/k$ for a star of $k$ nodes), after which redirection
shifts the attachment to the center of the star, thereby creating a perfect
star of $k+1$ nodes.  By this reasoning
\begin{equation}
\label{star}
S_N = \frac{2}{3}\times \frac{3}{4}\times\frac{4}{5}\times\cdots\times
\frac{N-2}{N-1}=\frac{2}{N-1}\,.
\end{equation}
The slower than exponential decay with $N$ of the star probability provides a
first clue that typical network realizations should be star like, as seen in
Fig.~\ref{examples}.  To make this surmise stronger, we compute the
probability to create a star graph with a single defect.  Such a network
arises by first building a perfect star of $k$ nodes, then making an
``error'' in which the new node attaches to the periphery of the star, and
finally building the rest of the star (Fig.~\ref{defect-star})

From Eq.~\eqref{star}, the probability to build a perfect star of $k$ nodes
is $\frac{2}{k-1}$.  A defect now occurs with probability $\frac{1}{k}$
because the new node must attach to the center of the star to create this
defect.  Finally, the probability that all remaining attachments occur to the
hub is
\begin{align}
\label{gamma}
\frac{k-3/2}{k+1}\times \frac{k-1/2}{k+2}\times\cdots\times
  \frac{N-5/2}{N}=\frac{\Gamma(k+1)}{\Gamma(k-3/2)}\,
\frac{\Gamma(N-3/2)}{\Gamma(N+1)}\,.
\end{align}
To understand each factor in the product, note that in a network of $n$ nodes
with a single defect, there is one hub, one nucleus node (of degree 2), one
leaf that attaches to the nucleus node, and $n-3$ leaves that attach to the
hub (Fig.~\ref{defect-star}).  To continue the star, the new node must either
select one of the $n-3$ leaves attached to the hub and then redirect to the
hub, or select the nucleus node and then redirect to the hub, which occurs
with probability $\frac{1}{n}\times \frac{1}{2}$.  The probability the new
node is redirected to the hub therefore is $(n-3+\frac{1}{2})/n$.

The probability to create a star of $N$ nodes with a single defect after $k$
nodes has been added, which we define as $S_{N,k}$, is
\begin{align}
\label{SNk}
S_{N,k}=  \frac{2}{k(k-1)}\,\frac{\Gamma(k+1)}{\Gamma(k-3/2)}\,
\frac{\Gamma(N-3/2)}{\Gamma(N+1)} = 2\,\frac{\Gamma(k-1)}{\Gamma(k-3/2)}\,
\frac{\Gamma(N-3/2)}{\Gamma(N+1)}\,.
\end{align}
Therefore the probability $S_N^{(1)}$ to build a star of $N$ nodes with a
single defect at any stage is given by
\begin{equation}
\label{S1:sum}
S_N^{(1)} = \sum_{k=3}^{N-2} S_{N,k} + \frac{2}{(N-1)(N-2)}\,,
\end{equation}
where the last term is the probability to create the defect after building a
perfect star of $N-1$ nodes.  Using \eqref{SNk} we compute the sum in
\eqref{S1:sum} and obtain~\cite{GKP89}
\begin{equation}
\label{S1:exact}
S_N^{(1)} = \frac{4}{3N} - \frac{2}{(N-1)(N-2)} + \frac{9}{N(N-1)(N-2)} - \frac{4}{3\sqrt{\pi}}\,\frac{\Gamma(N-3/2)}{\Gamma(N+1)}\,.
\end{equation}
Since dominant contribution to the sum comes from the terms with $k\gg 1$,
the leading behavior can be extracting by using the asymptotic,
\begin{equation*}
S_{N,k}=2\,\frac{\Gamma(k-1)}{\Gamma(k-3/2)}\,
\frac{\Gamma(N-3/2)}{\Gamma(N+1)}\simeq 2\,k^{1/2}\,N^{-5/2}\,,
\end{equation*}
for $k\gg 1$, and replacing summation in \eqref{S1:sum} by integration:
\begin{align}
\label{S1}
S_N^{(1)}\simeq\int^N \!\! S_{N,k}\, dk \simeq 2 N^{-5/2}\int^N \!k^{1/2}\,  dk 
\simeq \frac{4}{3N}\,.
\end{align}
We will use this procedure to show that multiple-defect stars arise with
roughly the same frequency as single-defect stars (\ref{app:defect}).

Comparing \eqref{star} and \eqref{S1} we see that a perfect star is $50\%$
more common than a single-defect star.  Naively, one would expect to find a
prescribed structure with a probability that is inversely proportional to the
total number of networks.  The latter grows factorially with $N$; e.g., the
number of labeled trees equals $N^{N-2}$~\cite{Cayley,O48,BOOK}.  By a
computation similar to \eqref{star}, the probability to build a linear graph
in the IR model equals $2/(N-1)!$ and thus agrees with naive expectations.
In contrast, perfect and slightly defective stars occur much more frequently
than naively expected.

More importantly, the above reasoning shows that star-like subgraphs will be
common in typical network realizations.  Consider such a structure, in which
the degree of the hub is $n$.  As will be shown in the next section, $n$
ranges from $a N$ to $b N$, where $0<a,b<1$.  Thus the probability that there
is a star-like module with the degree of the hub in this range is of the
order of
\begin{align}
\int_{aN}^{bN} \frac{dn}{n} = \ln \frac{b}{a}\,.
\end{align}
That is, with a non-zero (and scale independent) probability, there will be a
star-like structure whose degree is of the order of $N$, as observed in
Fig.~\ref{examples}.

\section{Maximal Degrees}
\label{sec:MD}

\begin{figure}[ht]
\centering
\includegraphics[width=0.45\textwidth]{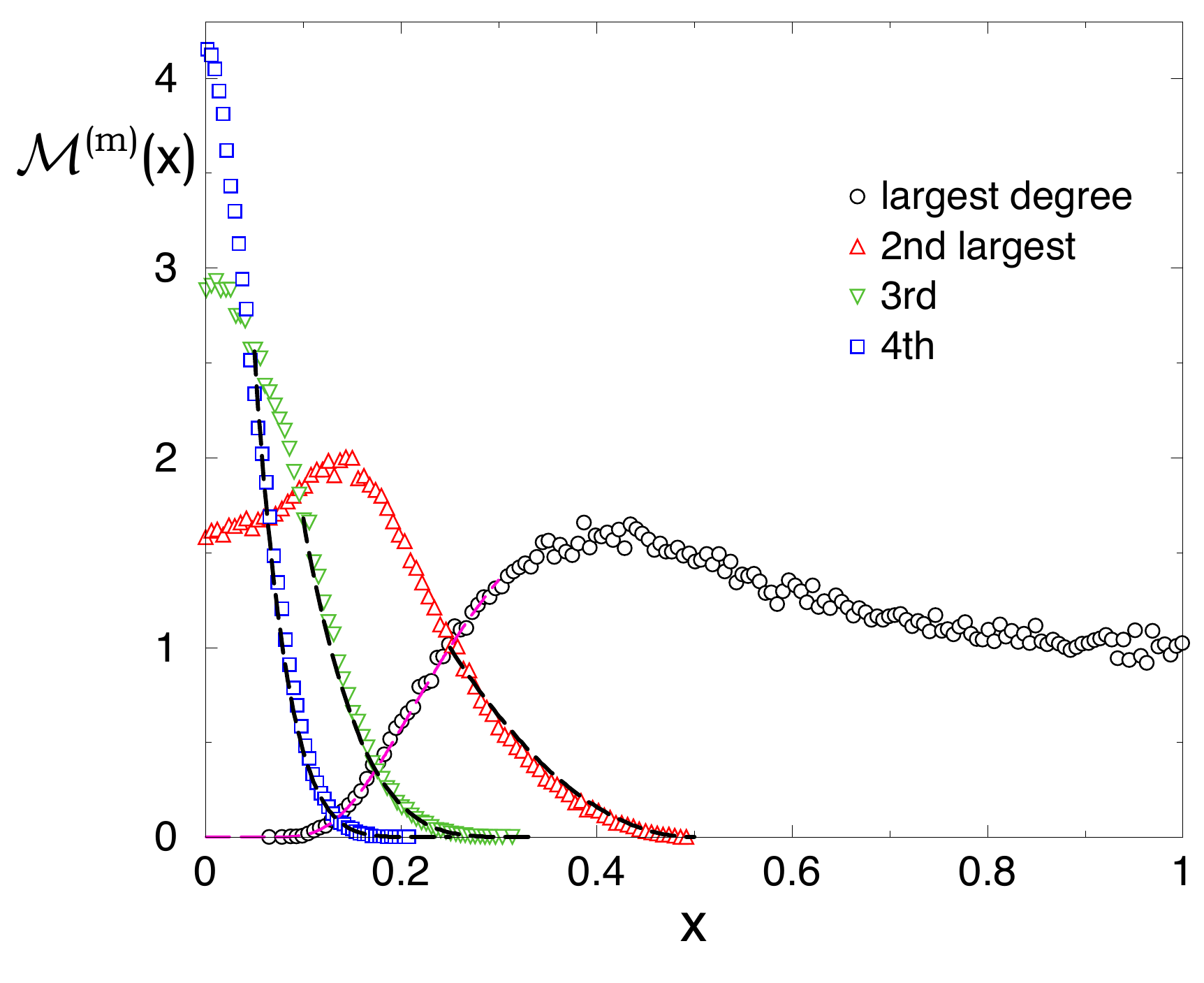}
\caption{The distributions $\mathcal{M}^{(m)}(x)$ versus normalized degree
  $x=k/N$ for $m\leq 4$ for $10^6$ realizations of networks with $N=10^5$.
  The initially normalized distributions for the $m^{\rm th}$ largest degree
  are divided by $m$, so that these rescaled distributions fit on the same
  plot.  The dashed line that follows the data for $\mathcal{M}^{(1)}$ is the
  empirical fit $-\frac{1}{2x}\ln(1/(4x))$, while the dashed lines that
  follow that data for $\mathcal{M}^{(2)}$, $\mathcal{M}^{(3)}$, and
  $\mathcal{M}^{(4)}$ are the predictions from Eq.~\eqref{Mh}, with the
  amplitudes 16, 600, and 40000 for the $2^{\rm nd}$, $3^{\rm rd}$, and
  $4^{\rm th}$ largest degree.}
\label{Fig:largest}
\end{figure}

Because macrohubs---nodes whose degree is a finite fraction of $N$---are at
the center of star-like graphs, they should occur at the same frequency as
stars.  This fact leads us to investigate the statistical properties of the
maximal network degree, $k_{\rm max}$, and also the $m^{\rm th}$ largest
degree $k_m$, in IR networks.  The value of $k_{\rm max}$ in the ensemble of
all networks of $N$ nodes is a random quantity that ranges between 2 and
$N-1$.  The smallest value $k_{\rm max} = 2$ arises for a linear graph, while
the largest value $k_{\rm max} = N-1$ arises for a star.  As
Fig.~\ref{Fig:largest} shows, $k_{\rm max}$, and indeed $k_m$ for any finite
$m$, scales linearly with $N$ in our IR model.  This scaling contrasts with
sparse networks, where $k_{\rm max}$ typically scales sublinearly with $N$.
Because there is a non-zero probability that the maximal degree is close to
$N$, macrohubs will be common in isotropic networks that grow by redirection;
a similar feature arises in enhanced redirection~\cite{GKR13}.

Let $M_N(k)$ denote the probability that the maximal degree in a network of
$N$ nodes equals $k$.  This largest degree is distributed over a wide range,
but is typically larger than $0.4N$ (Fig.~\ref{Fig:largest}).  It is
convenient to write this distribution in the scaling form
\begin{equation}
\label{scaling}
M_N(k)\to \frac{1}{N}\,\mathcal{M}^{(1)}(x), \qquad x = \frac{k}{N}
\end{equation}
as $k,N\to\infty$, with finite rescaled degree $x$.  The prefactor $N^{-1}$
imposes the normalization $\int_0^1 dx\,\mathcal{M}^{(1)}(x) = 1$.  Because
the distribution $\mathcal{M}^{(1)}(x)$ does not sharpen as $N$ increases,
moments of this distribution do not self-average~\cite{Der}; moreover, the
distribution is singular as $x\to 0$.  Similar singularities arise in a
variety of non-self-averaging processes, such as random maps, random walks,
spin glasses, and fragmentation processes~\cite{Der,DF,Higgs,FIK,DJM,KGB}.
In these systems, it was found that $\mathcal{M}^{(1)}(x)$ has an essential
singularity of the form $\exp[-x^{-1}\ln(1/x)]$ as $x\to 0$.  The same
singularity apparently occurs here.  Indeed, matching the exact result
$M_N(2)= 2/(N-1)!$ for the minimal possible degree with the scaled form
$\frac{1}{N}\,\mathcal{M}^{(1)}\left(\frac{2}{N}\right)$ gives
\begin{equation}
\label{essential}
\ln \mathcal{M}^{(1)} \sim -\frac{2}{x}\Big(\ln \frac{2}{x}-1\Big)\,,
\end{equation}
which qualitatively captures the small-$x$ behavior of
$\mathcal{M}^{(1)}(x)$.  However, we must be cautious in making this
connection because the scaled form is formally applicable when the rescaled
degree is finite, while we used $x=\frac{2}{N}\to 0$ in connecting the data
to the scaling form.  In fact, we find a good visual fit using
$\ln \mathcal{M}^{(1)}(x)\sim -\frac{1}{2x}\ln\big(\frac{1}{4x}\big)$.

More generally, the distributions $\mathcal{M}^{(m)}(x)$ for the $m^{\rm th}$
largest degree have support on $\big[0,\frac{1}{m}\big]$ and exhibit
power-law singularities as $x\to \frac{1}{m}$ from below
(Fig.~\ref{Fig:largest}).  We will derive this singular behavior in the next
section from the limiting behavior of the probability to find macrohubs of
specific topologies.

\section{Macrohubs}
\label{sec:H}

\begin{figure}[ht]
\centering
\includegraphics[width=0.3\textwidth]{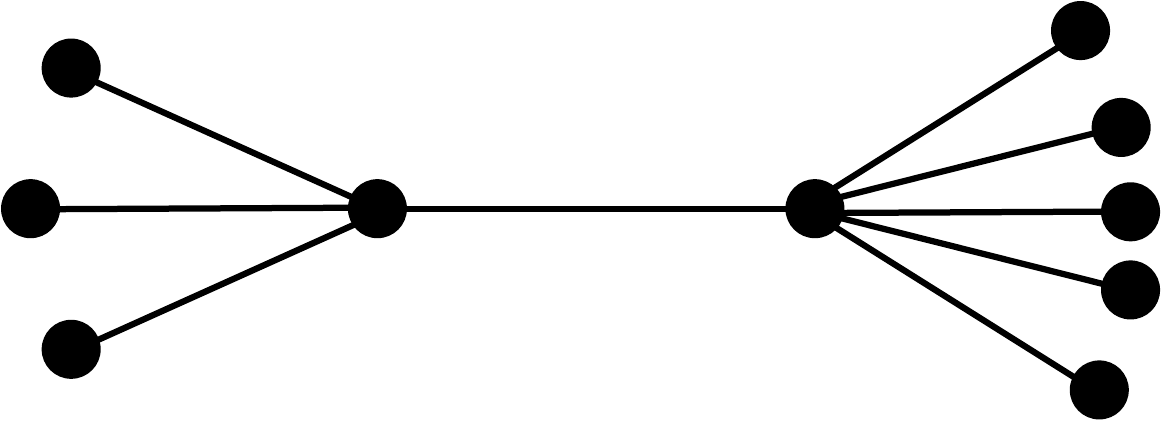}
\caption{The $(m,n)=(3,5)$ graph. }
\label{Fig:H35}
\end{figure}

As illustrated in Fig.~\ref{examples}, typical IR network realizations
contain multiple macrohubs.  To appreciate why such configurations are
common, let us examine the likelihood that there are exactly two connected
hubs, while all remaining nodes are leaves (see also Ref.~\cite{GKR13}).
Suppose that one hub is connected to $m$ leaves and the other to $n$ leaves,
leading to what we define as the $(m,n)$ graph (Fig.~\ref{Fig:H35}).  The hub
degrees are $m+1$ and $n+1$, respectively, and the total number of nodes is
$m+n+2$.  We term these nodes as hubs even if one of their degrees happens to
be small.

Let $H_{m,n}$ be the probability to build an $(m,n)$ graph.  Because, this
graph arises from $(m-1,n)$ and $(m,n-1)$ graphs, we can express $H_{m,n}$
through $H_{m-1,n}$ and $H_{m,n-1}$:
\begin{align}
\label{Hmn}
H_{m,n} &=  \frac{m-1+(n+1)^{-1}}{m+n+1}\, H_{m-1,n}
              +  \frac{n-1+(m+1)^{-1}}{m+n+1}\,H_{m,n-1}
\end{align}
For example, to build an $(m,n)$ graph from an $(m-1,n)$, the new node can
either select one of the $m-1$ leaves on the left of an $(m-1,n)$ graph and
redirect to the hub on the left.  The probability for this event is
$(m-1)/(m+n+1)$.  Additionally, the new node can select the right hub and
redirect to the left hub.  This event occurs with probability
$1/\big[(n+1)(m+n+1)\big]$.

We now solve \eqref{Hmn} for several relevant cases that help reveal the
star-like nature of typical IR network realizations.

\subsection{Two large hubs: $m,n\gg 1$}

When the hub degrees are both large, we treat $m$ and $n$ as continuous
variables and expand $H_{m-1,n}$ and $H_{m,n-1}$ in the Taylor series
\begin{equation*}
H_{m-1,n} = H - \frac{\partial H}{\partial m}\,, \quad
H_{m,n-1} = H - \frac{\partial H}{\partial n}\,,
\end{equation*}
where $H=H_{m,n}$.  Using this in \eqref{Hmn} gives
\begin{equation*}
m\,\frac{\partial H}{\partial m} + n\,\frac{\partial H}{\partial n}= - 3H\,.
\end{equation*}
The solution that satisfies the necessary symmetry requirement
$H_{m,n}=H_{n,m}$ is
\begin{equation}
\label{Hmn:large}
H_{m,n} = \frac{C_2}{(mn)^{3/2}}\,,
\end{equation}
where the amplitude $C_2$ is not computable within the continuum
approximation.

From \eqref{Hmn:large}, the probability for the
$\big(\frac{N}{2},\frac{N}{2}\big)$ graph scales as $N^{-3}$.  This also
gives the tail behavior of distribution of the second-largest degree, because
the probability for the $\big(\frac{N}{2},\frac{N}{2}\big)$ graph coincides
with the probability that the second-largest degree equals $N/2$.  Similar to
the distribution of the largest degree (Eq.~\eqref{scaling}), we anticipate
that the second-largest degree distribution has the scaling behavior
\begin{equation}
\label{scaling_2}
\frac{1}{N}\,\mathcal{M}^{(2)}(x), \qquad\text{with}\qquad x = \frac{k}{N}\leq \frac{1}{2}\,.
\end{equation}
This form is compatible with the above $N^{-3}$ probability for the
$\big(\frac{N}{2},\frac{N}{2}\big)$ graph if
\begin{equation}
\label{M2}
\mathcal{M}^{(2)}(x)\sim \left(\tfrac{1}{2}-x\right)^2 \quad\text{when}\quad  x\to \tfrac{1}{2}\,.
\end{equation}
Visually, this asymptotic behavior quantitatively agrees with simulation
results when an appropriate amplitude is chosen (Fig.~\ref{Fig:largest}).

\subsection{One large and one small hub: $m$ finite and $n\gg 1$}
 
\subsubsection{$m=1$}

Suppose that the degree of the smaller hub $m=1$.  This $(1,n)$ graph is also
just the single-defect star shown in Fig.~\ref{defect-star}.  The number of
nodes in such graph is $N=n+3$ and the largest degree is
$k_{\rm max} = n+1=N-2$.  To compute $H_{1,n}$, we must write the analog of
the recursion \eqref{Hmn} that applies for $m=1$.  This recursion is
\begin{equation}
\label{H1}
H_{1,n} = \frac{2}{(n+1)(n+2)} + \frac{n-\frac{1}{2}}{n+2}\,H_{1,n-1}\,.
\end{equation}
The first term on the right-hand side is the contribution that arises by
creating the $(1,n)$ graph from a perfect star with $n+2$ nodes.  To solve
\eqref{H1}, note that the homogeneous version of \eqref{H1} admits the
solution $\Gamma(n+\frac{1}{2})/\Gamma(n+3)$.  We use this solution as an
integrating factor
\begin{equation}
H_{1,n} = \frac{\Gamma(n+\frac{1}{2})}{\Gamma(n+3)}\,A_n\,,
\end{equation}
and substitute this form into \eqref{H1} to give
\begin{equation*}
A_{n} = A_{n-1}+2\,\frac{\Gamma(n+1)}{\Gamma(n+\frac{1}{2})}\,.
\end{equation*}
Because $H_{1,2}=\frac{5}{12}$ (see Fig.~\ref{enum}), the initial condition
for this recursion is $A_2={40}/(3\sqrt{\pi})$.  Thus~\cite{GKP89}
\begin{align}
\label{An}
A_{n} &= 2\sum_{j=3}^n \frac{\Gamma(j+1)}{\Gamma(j+\frac{1}{2})} +  \frac{40}{3\sqrt{\pi}} 
          = \frac{2}{3}\left[\frac{2\Gamma(n+3)- 3\Gamma(n+2)}{\Gamma(n+\frac{3}{2})} + \frac{4}{\sqrt{\pi}} \right]\,,
\end{align}
which leads to
\begin{equation}
\label{H1n}
H_{1,n} = \frac{2\Gamma(n+\frac{1}{2})}{3\Gamma(n+3)}\!
\left[\frac{2\Gamma(n+3)-3\Gamma(n+2)}{\Gamma(n+\frac{3}{2})} + \frac{4}{\sqrt{\pi}} \right]
\end{equation}
for $n\geq 2$.  The asymptotic behavior of \eqref{H1n} is
\begin{equation}
\label{H1n:asymp}
H_{1,n} \simeq \frac{U_1}{n}\,, \qquad U_1 = \frac{4}{3}\,,
\end{equation}
which coincides with the asymptotic probability for a single-defect star
given in Eq.~\eqref{S1}.  As another useful consequence of \eqref{H1n},
notice that a node of degree $N-2$ appears only in the $(1,N-3)$ graph.
Therefore $H_{1,N-3} = M_N(N-2)$, so that \eqref{H1n} also gives the
probability that the largest degree in a graph of $N$ nodes equals $N-2$ .

\subsubsection{$m=2$}

In analogy with \eqref{H1}, the recurrence for $H_{2,n}$ is
\begin{equation}
\label{H2}
H_{2,n} = \frac{n+2}{(n+1)(n+3)}\,H_{1,n} + \frac{n-\frac{2}{3}}{n+3}\,H_{2,n-1}\,.
\end{equation}
Using the homogeneous solution, one can again define the integrating factor
$H_{2,n} = \big[{\Gamma(n+\frac{1}{3})}/{\Gamma(n+4)}\big]B_n$, which reduces
\eqref{H2} to
\begin{equation*}
B_{n} = B_{n-1}+\frac{n+2}{n+1}\,\frac{\Gamma(n+\frac{1}{2})}{\Gamma(n+\frac{1}{3})}\,A_{n}\,,
\end{equation*}
with $A_n$ given by \eqref{An}.  However, instead of deriving the exact
solution to this equation, it is easier to extract the asymptotics by taking
the continuum limit of Eq.~\eqref{H2} to give
\begin{equation*}
\left(n\,\frac{d}{dn} + 3+\frac{2}{3}\right)H_{2,n} \simeq \frac{4}{3n} \,,
\end{equation*}
from which
\begin{equation}
\label{H2n:asymp}
H_{2,n} \simeq \frac{U_2}{n}\,, \qquad U_2=\frac{1}{2}\,.
\end{equation}
Note that this result also gives the probability for the 2-defect star shown
in Fig.~\ref{2defect-star}(b).

\subsection{$m=O(1)$}

More generally, for $m=O(1)$ and $n\gg 1$, we apply the continuum approach to
the large variable in \eqref{Hmn} and obtain
\begin{equation}
\label{H:diff}
\left(n\,\frac{\partial }{\partial n} + m+2-\frac{1}{m+1}\right)H_{m,n} =  H_{m-1,n} 
\end{equation}
Based on \eqref{H1n:asymp} and \eqref{H2n:asymp} we again expect that
\begin{equation}
\label{HU}
H_{m,n} = \frac{U_m}{n} 
\end{equation}
for $m=O(1)$ and $n\gg 1$.  Substituting \eqref{HU} into \eqref{H:diff}, the
amplitudes satisfy the recursion
\begin{equation*}
U_m = \left(m+1-\frac{1}{m+1}\right)^{-1}U_{m-1}\,,
\end{equation*}
from which 
\begin{equation}
\label{Um}
U_m = 2\prod_{j=1}^m \left(j+1-\frac{1}{j+1}\right)^{-1}= 4\,\frac{m+1}{(m+2)!}\,.
\end{equation}
As a postscript, note that $D_2\equiv \sum_{m\geq 1}U_m=2$, so that the
probability for all $(m,n)$ graphs with $m=O(1)$ and $n\gg 1$ is dominated by
the first two terms, for which $U_1+U_2=\frac{11}{6}$.

\subsection{More than two hubs}

\begin{figure}[ht]
\centering
\includegraphics[width=0.3\textwidth]{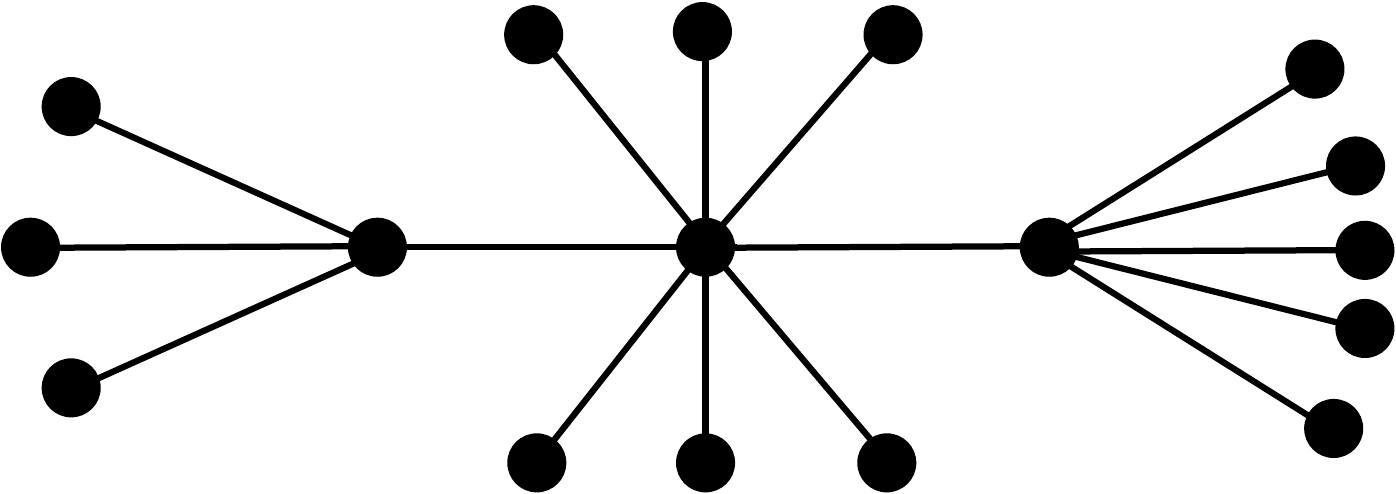}
\caption{The $(\ell,m,n)=(3,6,5)$ graph. }
\label{Fig:H365}
\end{figure}

To understand the general behavior, consider first the case of three
macrohubs.  Let $\ell,m,n$ denote the number of leaves connected to hubs of
degrees $\ell+1$, $m+2$, and $n+1$.  Note that the `central' hub with $n$
leaves is special because it is linked to both other hubs
(Fig.~\ref{Fig:H365}).  The total number of nodes is $\ell+m+n+3$.  Denote by
$H_{\ell,m,n}$ the probability to build such an $(\ell,m,n)$ graph.  This
graph can arise from $(\ell-1,m,n)$, $(\ell,m-1,n)$ and $(\ell,m,n-1)$
graphs, which occur with probabilities $H_{\ell-1,m,n}$, $H_{\ell,m-1,n}$ and
$H_{\ell,m,n-1}$.

These probabilities satisfy the recursion
\begin{align}
\label{Hmnl}
H_{\ell,m,n} & =  \frac{\ell-1+(m+2)^{-1}}{\ell+m+n+2}\, H_{\ell-1,m,n} 
                    +   \frac{m-1+(\ell+1)^{-1}+(n+1)^{-1}}{\ell+m+n+2}\,H_{\ell,m-1,n} \nonumber\\
                   & \hskip 2cm +  \frac{n-1+(m+2)^{-1}}{\ell+m+n+2}\,H_{\ell,m,n-1}\,.
\end{align}
Using continuum approach and assuming that all three hubs have large degrees,
i.e., $\ell,m,n\gg1$, we recast \eqref{Hmnl} into
\begin{equation*}
\ell\,\frac{\partial H}{\partial \ell} + m\,\frac{\partial H}{\partial m} + n\,\frac{\partial H}{\partial n}  + 5H = 0\,,
\end{equation*}
from which
\begin{equation}
\label{Hmnl:large}
H_{\ell,m,n} = \frac{C_3}{(\ell mn)^{5/3}}\,.
\end{equation}

Suppose that all three hubs are macroscopic, i.e., their degrees are linear
in the number of nodes: $(\ell,m,n) = N(a,b,c)$ with $a,b,c>0$ and $a+b+c=1$.
The probability $H_N(a,b,c)$ for such a three-hub network is
\begin{equation}
\label{H:abc}
H_N(a,b,c) = \frac{C_3 (abc)^{-5/3}}{N^5}
\end{equation}
Thus the probability for the $\big(\frac{N}{3},\frac{N}{3},\frac{N}{3}\big)$
graph scales as $N^{-5}$, which coincides with the probability that the
third-largest degree has the maximal possible size $\frac{N}{3}$.  This
third-largest degree also has the scaling behavior
$N^{-1}\mathcal{M}^{(3)}(x)$, which is compatible with the $N^{-5}$ extremal
behavior for the probability of three macrohubs when
\begin{equation}
\label{M3}
\mathcal{M}^{(3)}(x)\sim \left(\tfrac{1}{3}-x\right)^4 
\qquad\qquad  x\to \tfrac{1}{3}\,.
\end{equation}

Generally, the probability for a graph with $h$ macrohubs depends on the
nature of the links between these hubs when $h\geq 3$.  Nevertheless, if all
the macrohub degrees $m_j$ are large, the hub probability $H_{\bf m}$, with
${\bf m}=(m_1,\ldots,m_h)$, now satisfies
\begin{equation*}
\sum_{j=1}^h m_j\,\frac{\partial H}{\partial m_j}  + (2h-1)H = 0
\end{equation*}
from which
\begin{equation}
\label{H:large}
H_{\bf m} = C_h \prod_{j=1}^h m_j^{1/h-2}
\end{equation}
When all hubs are macroscopic, that is, $m_j=Na_j$, with $0<a_j<1$ and
$\sum_{j=1}^h a_j=1$, the network is realized with probability
\begin{equation}
\label{Hh}
H_N({\bf a}) = \frac{C_h}{N^{2h-1}}  \prod_{j=1}^h a_j^{1/h-2}\,,
\end{equation}
where ${\bf a}=(a_1,\ldots, a_h)$.  From \eqref{Hh}, the scaled distribution
of the $h^\text{th}$ largest-degree macrohub has the extremal behavior
\begin{equation}
\label{Mh}
\mathcal{M}^{(h)}(x)\sim \left(\tfrac{1}{h}-x\right)^{2h-2} 
\end{equation}
close to the maximal possible value $x\to \tfrac{1}{h}$.  The simulation data
shown in Fig.~\ref{Fig:largest} is consistent with these singular behaviors
for $h=2$, 3, and 4.

\section{Network Nucleus}
\label{sec:core} 

\subsection{Sublinear Growth}

\begin{figure}[ht]
\centerline{\includegraphics[width=0.425\textwidth]{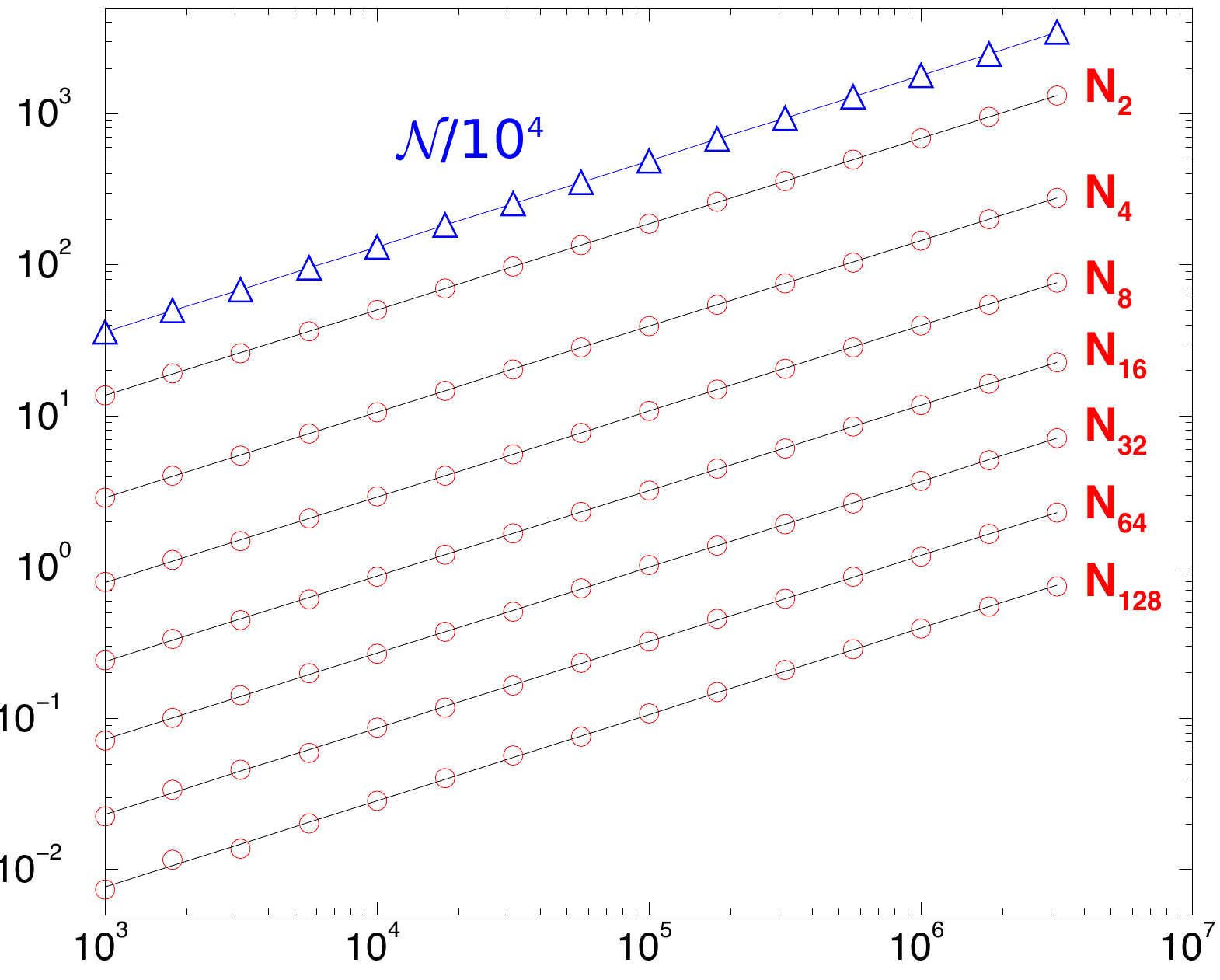}}
\caption{Dependence of $\mathcal{N}$ and $N_k$ versus $N$ for various $k$ values.}
\label{nk-vs-N}
\end{figure}

One of most enigmatic features of IR networks is that they consist almost
entirely of leaves, namely, nodes of degree one (Fig.~\ref{examples}).  Nodes
of degree greater than one constitute what we term the \emph{nucleus} of the
network.  Surprisingly, both the average number of nucleus nodes,
$\mathcal{N}= \sum_{k\ge 2}N_k$, and indeed the average number of nodes $N_k$ of any
fixed degree $k\geq 2$, grow sublinearly with $N$ (Fig.~\ref{nk-vs-N}):
\begin{equation}
\label{core}
\mathcal{N} \sim N^{\mu}\,,\qquad
N_k \sim N^{\mu}\,,
\end{equation}
with exponent $\mu\approx 0.566$.  The data are quite linear on the double
logarithmic scale of the figure and a linear fit gives a correlation
coefficient of 0.9999956.  In removing successive data points and performing
the same regression analysis, there is no systematic change to the slope,
which ranges between 0.5652 and 0.5673.  Our quoted exponent value of
$\mu=0.566$ therefore seems accurate to within 0.001.  Because of this
sublinear growth, the nucleus represents a vanishing fraction of the entire
network as $N\to\infty$, as is visually evident in Fig.~\ref{examples}.  This
behavior stands in stark contrast to that of sparse networks, where the
nucleus represents a finite fraction of the whole.

Starting with the sublinear scaling \eqref{core}, it is possible to generally
show~\cite{GKR13} that there is a power-law decay for the normalized degree
distribution, $c_k = {N_k}/{\mathcal{N}}$
\begin{equation}
\label{ck}
c_k\sim k^{-(1+\mu)} \qquad\qquad k\gg 1\,,
\end{equation}
with the degree distribution exponent $1+\mu$ less than 2.  Such an exponent
value can be shown to be mathematically inconsistent~\cite{GKR13} unless the
size of the nucleus grows sublinearly with $N$.

We now demonstrate that the nucleus of IR networks must grow sublinearly with
$N$.  Let $N_{k,\ell}$ be the number of nodes of degree $k$ that are
connected to $\ell$ leaves, and let $c_{k,\ell}={N_{k,\ell}}/{\mathcal{N}}$
be the density of such nodes.  We make the mild assumptions that $c_k$ and
$c_{k,\ell}$ are both independent of $N$ for $N\to\infty$.  Then the number
of nucleus nodes grows according to
\begin{equation}
\label{growth:core}
\frac{d\mathcal{N}}{dN}= \frac{\mathcal{N}}{N}
\sum_{k\geq 2}\sum_{\ell<k}c_{k,\ell}\, \frac{\ell}{k}\,.
\end{equation}
That is, the size of the nucleus increases by 1 only when a new node
initially selects a nucleus node and then redirects to a leaf.  A nucleus
node of degree $k$ that is attached to $\ell$ leaves is selected with
probability ${N_{k,\ell}}/{N}={\mathcal{N}}\,c_{k,\ell}/N$, and redirection
to a leaf occurs with probability $\frac{\ell}{k}$.  Nucleus nodes that are
not attached to leaves are characterized by $\ell=0$; attaching to these
nodes therefore does not affect the nucleus size.

If the densities $c_{k,\ell}$ are independent of $N$ as $N\to\infty$, then
\eqref{growth:core} implies that
\begin{equation}
\label{nu:sum}
\mu =  \sum_{k\geq 2}\sum_{\ell<k} c_{k,\ell}\, \frac{\ell}{k}\,.
\end{equation}
We now obtain the strict upper bound $\mu < 1$ by replacing $\ell$ by its
largest possible value, which is $k-1$, in the sum in \eqref{nu:sum}.  We
also exploit the two obvious sum rules, $\sum_{\ell<k} c_{k,\ell}=c_k$ and
$\sum_{k\geq 2} c_k=1$ to give
\begin{align*}
\sum_{k\geq 2}\sum_{\ell<k} c_{k,\ell}\, \frac{\ell}{k}  \leq  
\sum_{k\geq 2} c_k \, \frac{k-1}{k}= 1 - \sum_{k\geq 2} k^{-1} c_k < 1\,.
\end{align*}
We conclude that $\mu<1$, which gives the fundamental result that the nucleus
grows sublinearly with $N$.

\subsection{Anomalously small nucleus}
\label{subsec:core}

We previously showed there is an anomalously large probability, of order
$1/N$, to generate star-like graphs that necessarily have a small nucleus.
Therefore we anticipate that the probability to generate a graph with a
nucleus whose size is a finite number will also be proportional to $1/N$.  We
therefore focus on the probability that the nucleus has a finite size $h$:
\begin{equation}
P_h(N) \equiv \text{Prob}\big[\big|\mathcal{N}\big| = h\big]\,.
\end{equation}
Equivalently, this is the same as the probability that there are $h$
macrohubs in the system.  We already know the probability for the nucleus to
consist of a single macrohub,
\begin{equation*}
\label{P1}
P_1(N) \equiv S_N = \frac{2}{N-1}\,,
\end{equation*}
because this is the same as the probability to create a star graph.

Consider the probability that there two hubs.  Let us first suppose that the
degrees of both hubs are large; without loss of generality, we set $m\leq n$.
Summing over all partitions of the two hub degrees, a graph with $N$ nodes
has exactly 2 hubs with probability
\begin{equation}
\label{P2N:def}
P_2(N) = \sum_{m=1}^{\lfloor N/2\rfloor - 1} H_{m,N-2-m}\,.
\end{equation}
When both hubs are macroscopic, we previously derived in
Eq.~\eqref{Hmn:large} that $H_{m,n}\simeq N^{-3}$.  Moreover, the number of
such contributions to the above sum is of the order of $N$.  Hence the
overall contribution to this sum from macroscopic hubs scales as $N^{-2}$.

Thus the dominant contribution to the sum comes from terms with small $m$.
We therefore use the result from Eq.~\eqref{HU} that $H_{m,n}=U_m n^{-1}$ for
$m\sim O(1)$ and $n$ large and substitute into \eqref{P2N:def} to obtain
\begin{equation}
\label{P2N}
P_2(N) \simeq \frac{D_2}{N} =\frac{2}{N}\qquad N\to\infty\,.
\end{equation}
with $D_2$ defined immediately after Eq.~\eqref{Um}.  Therefore the
probabilities to generate either one hub---a star network---or two hubs are
asymptotically of the same order.

Based on these results, we anticipate that
\begin{equation}
\label{Ph}
P_h(N) \simeq \frac{D_h}{N}
\end{equation}
for arbitrary $h$.  In analogy with the discussion of two hubs, to justify
\eqref{Ph} it is necessary to determine $H_{\bf m}$ near the ``corner''
values of {\bf m}, where all hub degrees, apart from one, are small; the
contribution from the cases where all hubs have macroscopic degrees is
negligible.  As a first step, we analyze two illustrative cases with three
hubs in which one of them is macroscopic in \ref{app:multiple}.  Namely, we
show that the probabilities for the graphs $H_{1,m,1}$ and $H_{\ell,0,1}$
(see Eqs.~\eqref{1H1:sol} and \eqref{H01:sol}) are indeed proportional to
$N^{-1}$.

\section{Multiple Linking}
\label{sec:book}

We now extend IR networks in which a new node makes more than one link to the
network.  For sparse networks, this modification affects only the amplitude
of the degree distribution.  For example, for linear preferential attachment
in which the new node makes $m$ independent links to the network, the number
of nodes of degree $k$ asymptotically scales as (see e.g.,~\cite{KR01,N10})
\begin{equation*}
N_k \simeq \frac{2m(m+1) N}{k^3}\,.
\end{equation*}
That is, the exponent of the degree distribution does not depend on $m$, the
out degree of the new node.  In isotropic redirection, however, the degree
$m$ of the newly introduced node materially affects the degree distribution.

\begin{figure}[ht]
\vspace*{-1.55cm}
\hspace*{0.22cm}
\centerline{\subfigure[]{\includegraphics[width=0.275\textwidth]{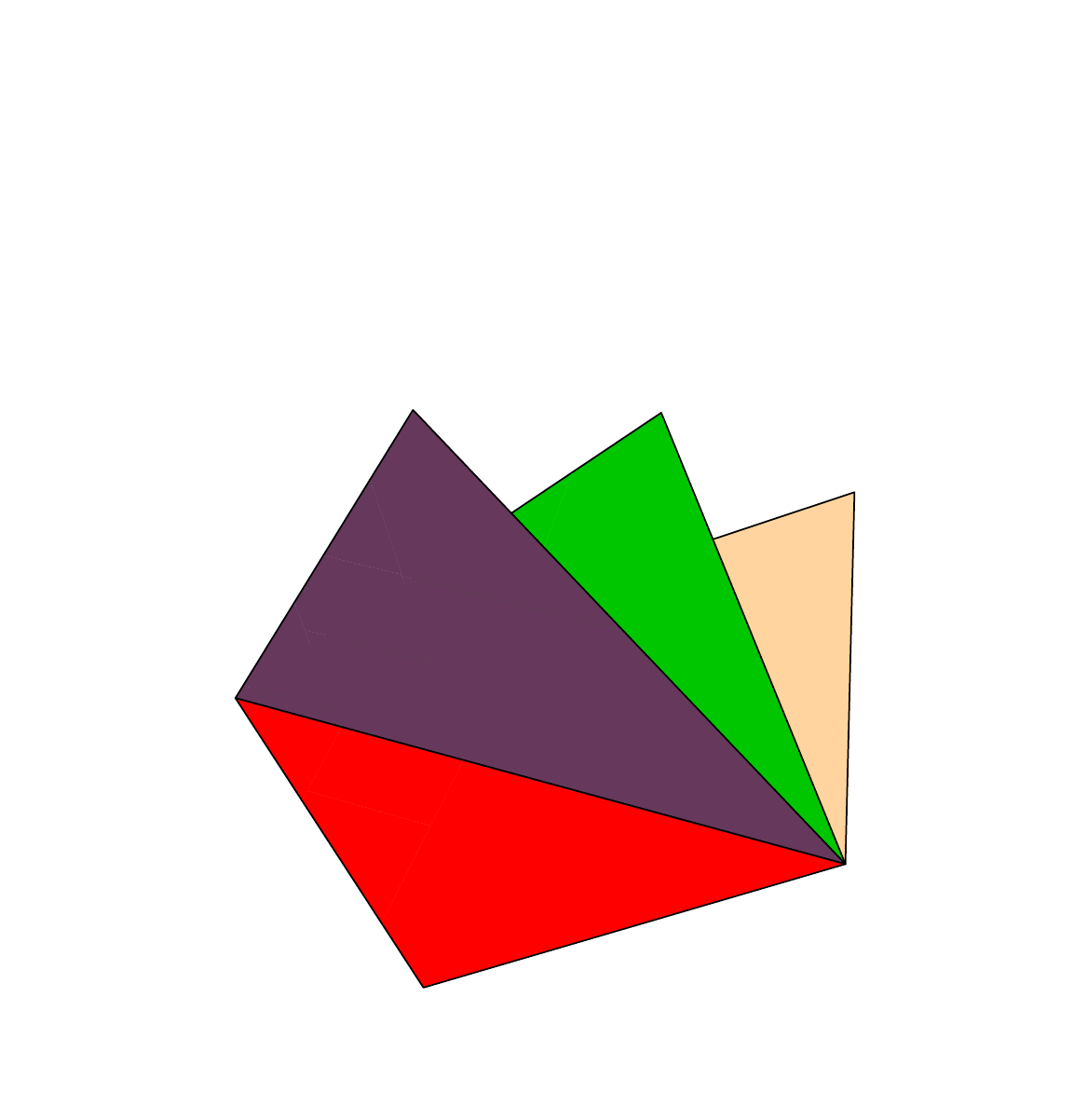}}
\hspace*{0.5cm}
\subfigure[]{\includegraphics[width=0.35\textwidth]{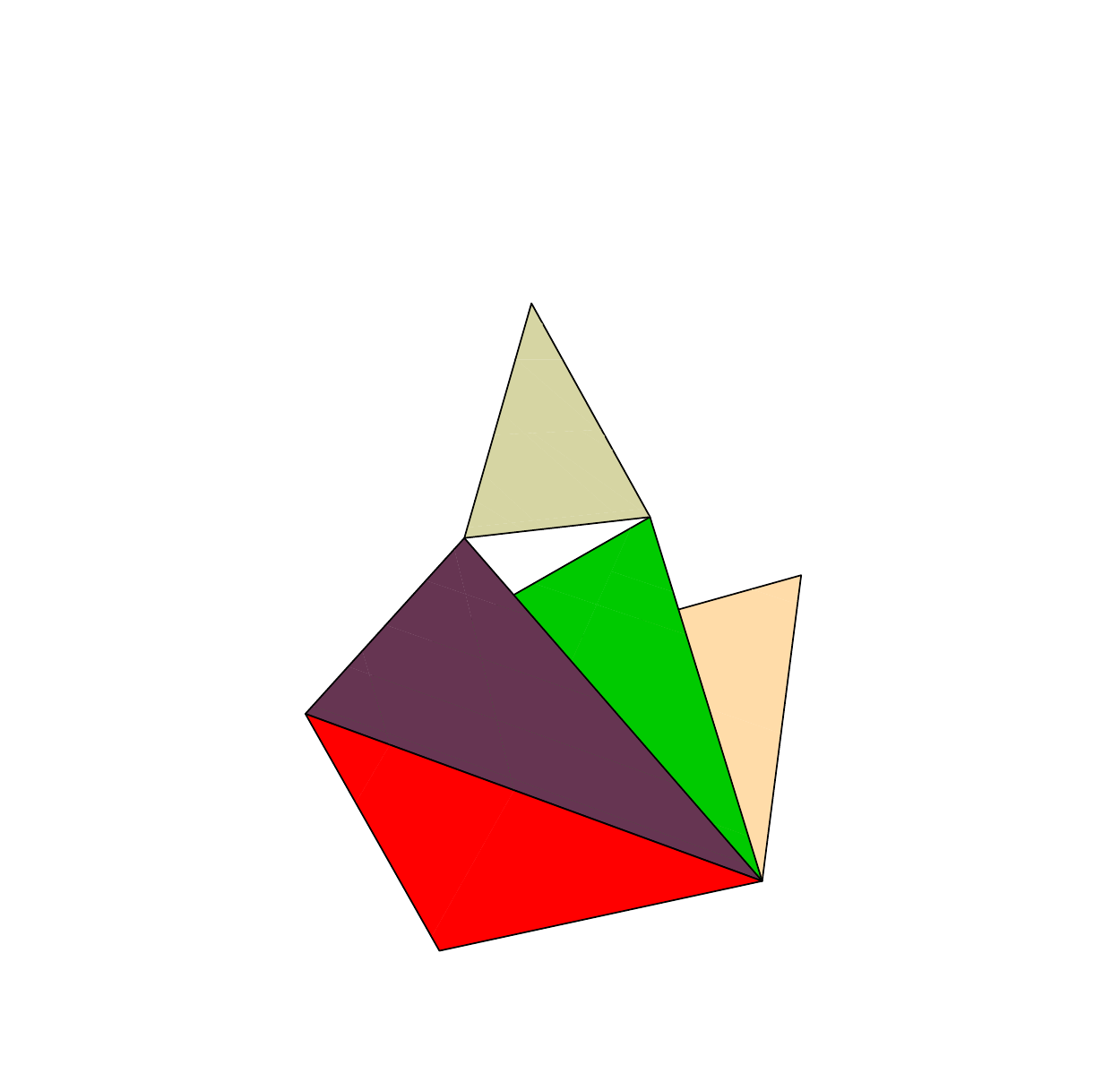}}
\subfigure[]{\includegraphics[width=0.375\textwidth]{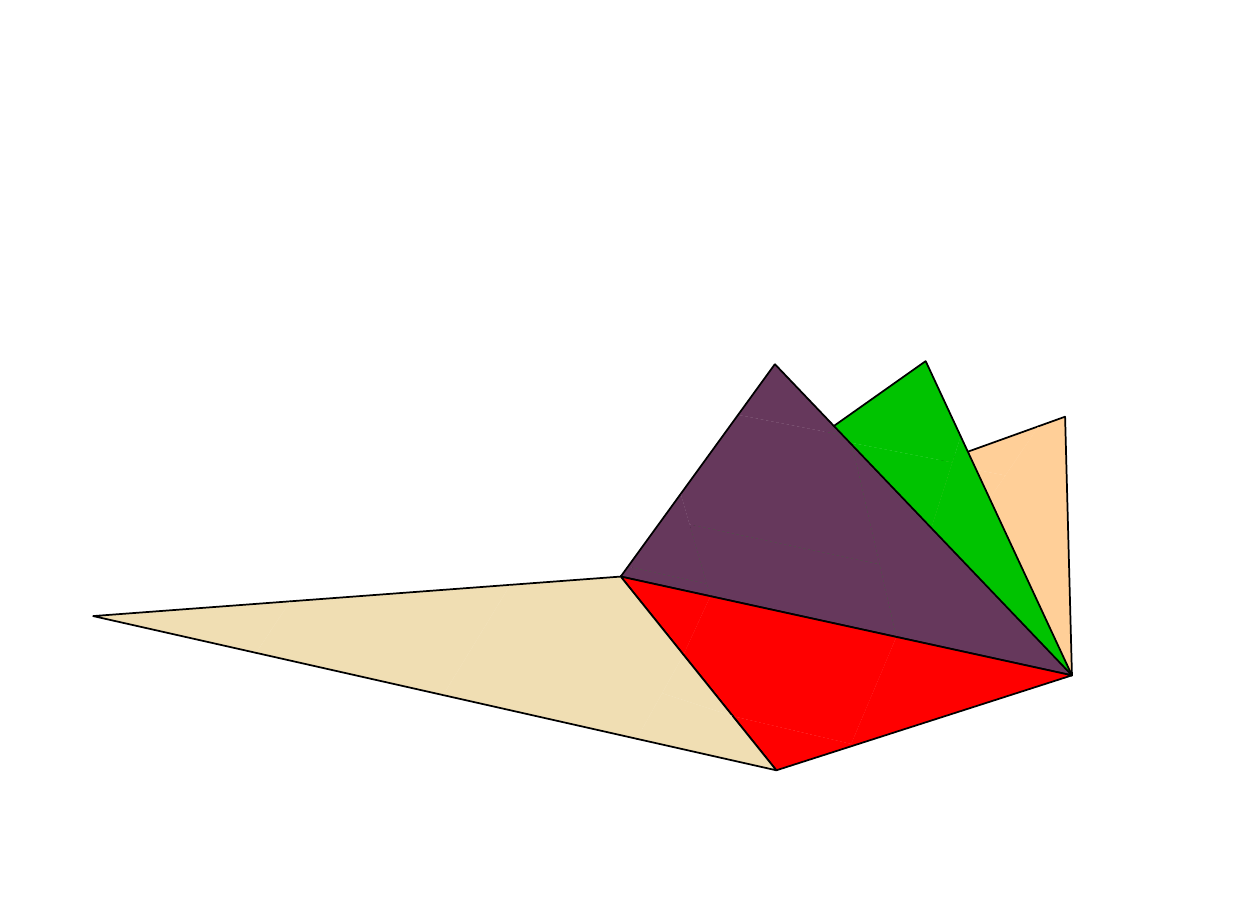}}}
\caption{(a) Illustration of a book with $M=4$ pages. (b) A defect in which
  the new node links to two lowest-degree nodes after a book of $M=6$ pages
  is created.  (c) A defect in which the new node links to a node of the
  lowest degree and a node of the highest degree.}
\label{Fig:pix}  
\end{figure}

There are two natural ways to implement multiple linking: (a) the new node
selects a provisional target at random and attaches to $m$ randomly selected
but distinct neighbors of this single target, or (b) the new node selects $m$
provisional random targets and attaches to a random neighbor of each of these
targets.  Let us first consider rule (a) with $m=2$; here, it is convenient
to take the initial condition as a triangle.  We may visualize the growth
process as an effective triangle being created each time a new node and two
new links are introduced (Fig.~\ref{Fig:pix}).  However, we emphasize that we
are generating a random graph, and not a set of random triangular surfaces.
(Random triangular surfaces and, more generally, random simplicial complexes
are studied, e.g., in \cite{T1,T2,T3,T:Dima,T4,T5}).

More generally, we study networks where each new node makes $m$ links to the
network; here, it is convenient to take the initial condition as a complete
graph of $m+1$ nodes.  Here, we define the nucleus of the network as the set
of nodes with degree $k>m$.  Once again the average number of nucleus nodes
$\mathcal{N}$, as well as the average number of nodes of any fixed degree
$N_k$, with $k>m$, both scale in the same way and sublinearly with $N$:
$\mathcal{N}\sim N^\mu$ and $N_k\sim N^\mu$.  The salient feature is that
exponent $\mu$ is non-universal with respect to $m$ (and also to the two
linking rules, (a) and (b)).  For the first few cases, the exponent values
are:
\begin{equation}
\mu =
\begin{cases}
0.74 & \qquad  \text{1 target, $m=2$ links to its neighbors}\\[-0.125cm]
0.83 & \qquad  \text{1 target, $m=3$ links to its neighbors}\\[-0.125cm]
0.88 & \qquad  \text{1 target, $m=4$ links to its neighbors}\\
0.83 & \qquad   \text{link to neighbors of $m=2$ targets}\\[-0.125cm]
0.93 & \qquad   \text{link to neighbors of $m=3$ targets}\\[-0.125cm]
0.97 & \qquad   \text{link to neighbors of $m=4$ targets}
\end{cases}
\end{equation}
As $m$ increases the scaling of $\mathcal{N}$ and $N_k$ gradually approaches
linearity in $N$.  We note that our values for the degree distribution
exponent $\nu=1+\mu$ are close to those reported in~\cite{MPP13} for various
Wikipedia pages where multiple linking is significant.

In analogy with star graphs when the new node make a single link, we also
examine the corresponding extremal graphs when the new node makes $m>1$
links.  For $m=2$ and rule (a), the analog of a star graph is a \emph{book}
(Fig.~\ref{Fig:pix}(a)).  A book with $N$ nodes has $N-2$ triangular pages
that all share a common link that acts as the binding between the two highest
degree nodes.  (In Ref.~\cite{KK:open} this graph was called an open book.)~
The probability $B_N$ to build a book of $N$ nodes is obtained by iterating
the recursion $B_{N+1}=\frac{N-2}{N}B_N$.  Starting with $B_4=1$, we obtain
\begin{equation}
\label{book}
B_N = \frac{6}{(N-1)(N-2)}  \qquad\quad N\geq 4\,.
\end{equation}

Consider now books with one defect, as illustrated in Fig.~\ref{Fig:pix}(b).
To compute the probability to create a single-defect book, one first
generates a book of $M$ nodes, and then make an error by selecting one of the
highest degree nodes as the target and thereby link to two nodes of the
lowest degree (Fig.~\ref{Fig:pix}(b)).  All subsequent growth steps continue
to build the book without any additional errors.  This configuration occurs
with probability
\begin{equation}
\label{BMN}
B(M,N)=B_M\,\frac{2}{M}\,\frac{M-3}{M-1}\prod_{k=M+1}^{N-1}\frac{k-\frac{13}{3}}{k}\,.
\end{equation}
Here $B_M$ accounts for generating a book with $M$ nodes, the factor
$\frac{2}{M}$ accounts for then ``erroneously'' choosing one of the two
highest degree nodes, and the factor
$\frac{M-3}{M-1}=\binom{M-2}{2}/\binom{M-1}{2}$ accounts for the linking to
the lowest degree nodes.  The probability for the remaining attachments to
occur without any errors is $(k-\frac{13}{3})/k$ when the total number of
nodes equals $k$.  Writing the product in terms of gamma functions gives
\begin{align}
\label{BMN-asymp}
B(M,N) &= 12\,\frac{(M\!-\!3)\,\Gamma(M\!-\!2) }{(M\!-\!1)\,\Gamma(M\!-\!\frac{10}{3})}\,\frac{\Gamma(N\!-\!\frac{13}{3})}{\Gamma(N)} \simeq 12\,M^{4/3}\,N^{-13/3}\qquad M\to\infty\,.
\end{align}
Thus the probability to create a network with a single defect of the type
illustrated in Fig.~\ref{Fig:pix}(b) is given by
\begin{equation}
\label{bN}
B'_N=\!\sum_{4\leq M\leq N-1} \!B(M,N) \simeq \frac{36}{7}\frac{1}{N^2}\qquad N\to\infty\,.
\end{equation}

Analogously, we can compute the probability to create a book with a single
error of the type shown in Fig.~\ref{Fig:pix}(c).  The probability to create
this defect after the network contains $M$ nodes is
\begin{align}
\label{AMN}
A(M,N)&=B_M\,\frac{2}{M}\,\frac{2}{M-1}\prod_{k=M+1}^{N-1}\frac{k-\frac{11}{3}}{k}
= 24\,\frac{\Gamma(M-2)
        }{(M-1)\,\Gamma(M-\frac{8}{3})}\,\frac{\Gamma(N-\frac{11}{3})}{\Gamma(N)}\,,\nonumber \\
&\simeq 24\,M^{-1/3}\,N^{-11/3}\qquad M\to\infty\,.
\end{align}
The total probability $A'_N=\sum_{4\leq M\leq N-1} A(M,N)$ to have one defect
of this type therefore scales as
\begin{equation}
\label{aN}
A_N'\simeq \frac{36}{N^3}\,.
\end{equation}
Thus defects of the type in Fig.~\ref{Fig:pix}(b) are more common than those
shown in~\ref{Fig:pix}(c).

\section{Discussion}

We introduced a parameter-free network growth mechanism---isotropic
redirection (IR)---that represents a minimalist extension of the classic
random recursive tree (RRT).  In the RRT, new nodes connect to an existing
network one by one.  Each new node selects and connects to a target node in
the existing network that is chosen uniformly at random.  In our IR model,
each new node again selects a random target node but then connects to one of
its neighbors.

In spite of the homogeneity of this simple growth rule, highly modular
networks emerge that contain multiple macrohubs---nodes whose degrees are
macroscopic (Fig.~\ref{examples}).  Visually, these networks share many
features with multiplex networks; the latter are comprised of well resolved
individual networks that are weakly interconnected.  It is remarkable that
the modular configurations characteristic of multiplex networks arise
essentially for free in our IR model.

A striking feature of network realizations in our IR model is that star-like
structures are quite common.  Naively, one might have anticipated that the
probability for the occurrence of these extremal configurations would be
exponentially small in $N$.  By probabilistic reasoning, we showed that the
likelihood of stars and near-perfect stars are both proportional to $N^{-1}$.
This anomalous probability results from the inherent amplification of the
redirection process as the degree of a central node starts to ``run away''
from the typical network degree.  Therefore star-like structures whose
central degree is of the order of $N$ occur with a non-zero probability.

An outstanding theoretical challenge is to determine the exponent $\mu$ that
characterizes the sublinear scaling of the average size $\mathcal{N}$ of the
``nucleus'' of the network.  Numerically, we found $\mathcal{N}\sim N^\mu$,
with $\mu\approx 0.566$.  Thus the nucleus represents a vanishingly small
fraction of the entire network.  This behavior again starkly contrasts with
sparse networks, where the size of the nucleus scales linearly with $N$.  If
the new node makes $m>1$ connections to the network then many of the
anomalous features observed for the case $m=1$ still arise, but the scaling
of nucleus size on $N$ appears to approach linearity as $m$ increases.

We thank D. ben-Avraham for discussions and collaboration on the IR model for
the general case of $r<1$.  We are grateful to G. S. Redner for his
assistance in translating our FORTRAN codes into C++ to take advantage of
dynamic memory allocation, a helpful feature for growing undirected networks.
We also acknowledge support from grants DMR-1608211 and DMR-1623243 from the
National Science Foundation, and by the John Templeton Foundation (SR).

\appendix

\section{Stars with Two or More Defects}
\label{app:defect}
 
We extend the approach given in Sec.~\ref{sec:star} to compute the
probability for a star with two or more defects.  We will show that
multiple-defect stars are more common than single-defect stars, thus
providing more evidence that typical network configurations contain many
star-like subgraphs.  To create a start with two defects, the following must
occur:
\begin{enumerate}
\setlength\itemsep{-0.35ex}
\item First build a perfect star of $k$ nodes.
\item Make an error when the next node is introduced.
\item Then continue attaching to the hub until a single-defect star of $\ell$
  nodes is made.
\item Make a second error when the next node is introduced.
\item Then continue attaching to the hub until a two-defect star of $N$ nodes is
  made.
\end{enumerate}

\begin{figure}[ht]
\centerline{\includegraphics[width=0.7\textwidth]{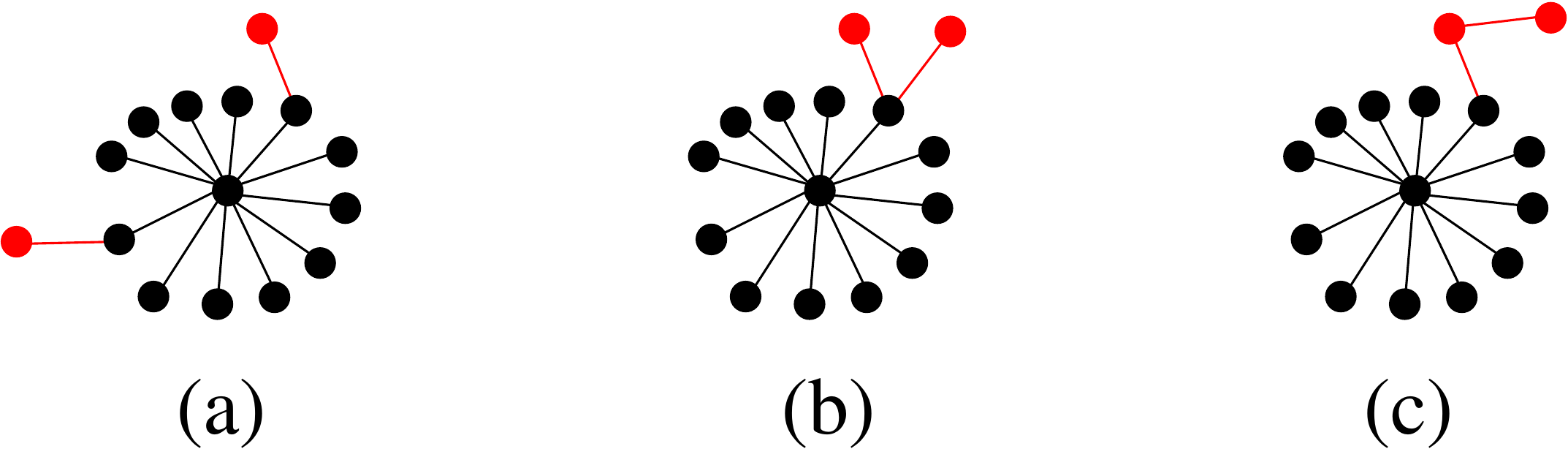}}
\caption{The three types of stars with 2 defects}
\label{2defect-star}
\end{figure}

There are three distinct types of 2-defect stars (Fig.~\ref{2defect-star}),
and we now calculate the probability to create each of them.  For each 
type, this probability generically has the asymptotic form
\begin{align}
\label{SNkl}
  S_{N,k,\ell}= \frac{2}{(k-1)}\,d_1(k)\, 
  \prod_{k+1}^\ell\Big(1-\frac{a_1}{n}\Big)\,\,d_2(\ell)\,
  \prod_{\ell+1}^N\Big(1-\frac{a_2}{n}\Big)\,.
\end{align}
The first factor is the probability to create a perfect star of $k$ nodes.
The second factor, $d_1(k)=\frac{1}{k}$, is the probability to create the
first defect.  The next factor is the probability to add new nodes to the
network without creating another defect; this probability was written in
Eq.~\eqref{gamma} with $a_1=\frac{5}{2}$.  The factor $d_2(\ell)$ is the
probability to create the second defect when the network contains $\ell$
nodes.  The last product gives the probability to build the network to its
final state without any additional defects.  Both $d_2(\ell)$ and $a_2$ depend
on the topology of the 2-defect star that is created.

Before specifying $d_2(\ell)$ and $a_2$, we first determine the asymptotic
behavior of the products in \eqref{SNkl}.  While we can write them in terms
of gamma functions, the following shortcut suffices for the asymptotic
behavior.  Generically, we write the products in Eq.~\eqref{SNkl} as
\begin{align}
\prod_{n=k}^\ell\Big(1-\frac{a_1}{n}\Big)& =
\exp\Big[\sum_{n=k}^\ell \ln\Big(1-\frac{a_1}{n}\Big)\Big]
\simeq \exp\Big[-\int_k^\ell \frac{a_1}{n}\,dn \Big]\simeq \left(\frac{k}{\ell}\right)^{a_1}\,.
\end{align}

We now determine $d_2(\ell)$ and $a_2$ for the distinct 2-defect stars in
Fig.~\ref{2defect-star}(a), (b), and (c).  For case (a), there is one hub,
one core node, one leaf attached to the nucleus node, and $\ell-3$ leaves
attached to the hub.  By enumerating all relevant states, the probability to
create a second defect is
\begin{equation*}
d_2(\ell) = \frac{1}{\ell}\,\frac{\ell-3}{\ell-2}\simeq \frac{1}{\ell}\,.
\end{equation*}
Once a second defect is created, a network of $n$ nodes consists of a hub,
two nucleus nodes, two leaves attached to nucleus nodes, and $n-5$ leaves attached
to the hub.  If the new node selects one of these $n-5$ leaves, then
attachment to the hub occurs.  If the new node selects one of the two nucleus
nodes, then with probability 1/2, redirection to the hub occurs.  Thus the
probability that a new node attaches to the hub is
$\big[n-5+2\times (1/2)\big]/n = 1-\frac{4}{n}$, so that
\begin{equation*}
a_2=4\,.
\end{equation*}
By enumerations in the same spirit, the results  for cases (b) and (c) are
\begin{align}
d_2(\ell)&= \frac{1}{\ell}\Big(1+\frac{1}{\ell-2}\Big)\simeq \frac{1}{\ell}\,,\qquad
a_2=\frac{11}{3}\,,\nonumber\\[0.3cm]
d_2(\ell)&= \frac{1}{2\ell}\,,\hspace{3.8cm}
a_2=\frac{7}{2}\,.\nonumber
\end{align}

Substituting these in Eq.~\eqref{SNkl}, integrating over the possible values
of $k$ and $\ell$, the probability to create an $N$ node 2-defect star of
type (a) is given by
\begin{align}
\label{SN2}
S_N^{(2)}\equiv \sum S_{N,k,\ell}&\simeq \int_1^N dk 
\int_k^N d\ell \,d_2(\ell)\,S_{N,k,\ell}\,,\nonumber \\
&\simeq\frac{2}{k^2}\,\int_1^N dk \int_k^N \left(\frac{k}{\ell}\right)^{a_1}
\frac{1}{\ell}\left(\frac{\ell}{N}\right)^{a_2}\,,\nonumber\\
&\simeq \frac{2}{N}\frac{1}{(a_1-1)(a_2-1)}= \frac{4}{9N}\,.
\end{align}
Note that this type of defective star can also be viewed as the graph of type
$(1,m,1)$, whose probability is determined independently in the next section.
The result \eqref{SN2} thus coincides with \eqref{1H1:sol}, the probability
to create a $(1,m,1)$ graph.  By similar calculations, the probability to
create 2-defect stars of types (b) and (c) are $\frac{1}{2N}$ and
$\frac{4}{15N}$, respectively.  The former reproduces \eqref{H2n:asymp},
while the latter reproduces \eqref{H01:sol}, which will also derived in the
next section.

\section{Multiple Hubs}
\label{app:multiple}

To help understand the behavior of the probability to generate a small number
of macrohubs, we compute the probability for specific networks with three
hubs when two hub degrees are small.  The enumerative procedure is
straightforward, albeit a bit tedious.  As a first example, consider
$H_{1,m,1}$.  In this case the analog of the recursion \eqref{Hmnl}, for the
case where the first and third arguments are small, is
\begin{equation}
\label{1H1}
H_{1,m,1} = \frac{m+1}{(m+2)(m+4)}\,H_{1,m+1} +  \frac{m}{m+4}\,H_{1,m-1,1}\,,
\end{equation}
with $H_{1,m+1}$ determined by \eqref{H1n}.  While this recurrence is
soluble, it suffices to use the continuum approach to determine the
asymptotic behavior.  Using \eqref{H1n:asymp}, we recast \eqref{1H1} into 
\begin{equation}
\label{1H1:ODE}
 \left(m\,\frac{d}{dm} + 4\right)H_{1,m,1} \simeq \frac{4}{3m} \,,
\end{equation}
from which
\begin{equation}
\label{1H1:sol}
H_{1,m,1} \simeq \frac{4}{9m} \,.
\end{equation}
This result coincides with \eqref{SN2}, as it must.

Another illustrative example is when the degree of the central hub is the
smallest; the simplest such example is the $(\ell,0,1)$ graph.  Here the
central hub is not linked to any leaf, but has degree 2.  For this limiting
case, the recursion for the number of such graphs is
\begin{equation}
\label{H01}
H_{\ell,0,1} = \frac{1}{2(\ell+3)}\,H_{1,\ell} + \frac{\ell-\frac{1}{2}}{\ell+3}\,H_{\ell-1,0,1}
\end{equation}
To obtain the asymptotic behavior, we use \eqref{H1n:asymp} and again take
the continuum limit to recast \eqref{H01} into
\begin{equation}
\label{H01:ODE}
 \left(\ell\,\frac{d}{d\ell} + \frac{7}{2}\right)H_{\ell,0,1} \simeq \frac{2}{3\ell} \,,
\end{equation}
from which
\begin{equation}
\label{H01:sol}
H_{\ell,0,1} \simeq \frac{4}{15\ell} \,.
\end{equation}

We have thus identified three-hub configurations whose occurrence
probabilities in the ensemble of networks of $N$ nodes is proportional to
$N^{-1}$.  By extending this reasoning, we anticipate that the probability to
find $N$-node networks of $h$ hubs in the full ensemble will also be
proportional to $N^{-1}$.

\end{document}